\documentclass[11pt]{article}

\usepackage{cmap}
\usepackage[T1]{fontenc}
\usepackage[margin=2.2cm]{geometry}
\usepackage{authblk}
\usepackage[numbers]{natbib}
\usepackage{booktabs}
\usepackage{longtable}
\usepackage{array}
\usepackage{amsmath,amssymb}
\usepackage{xcolor}
\usepackage{graphicx}
\usepackage{siunitx}
\usepackage{enumitem}
\usepackage[hidelinks]{hyperref}
\usepackage{tikz}
\usetikzlibrary{shapes.geometric,arrows.meta,positioning,fit,backgrounds,calc}

\definecolor{amBlue}{HTML}{185FA5}
\definecolor{amBlueF}{HTML}{E6F1FB}
\definecolor{amTeal}{HTML}{0F6E56}
\definecolor{amTealF}{HTML}{E1F5EE}
\definecolor{amCoral}{HTML}{993C1D}
\definecolor{amCoralF}{HTML}{FAECE7}
\definecolor{amAmber}{HTML}{854F0B}
\definecolor{amAmberF}{HTML}{FAEEDA}
\definecolor{amGray}{HTML}{5F5E5A}
\definecolor{amGrayF}{HTML}{F1EFE8}
\definecolor{amRed}{HTML}{A32D2D}
\definecolor{amRedF}{HTML}{FCEBEB}
\definecolor{amPurple}{HTML}{534AB7}

\tikzset{
  ucase/.style   ={ellipse, draw=amBlue, fill=amBlueF, line width=0.4pt,
                    align=center, font=\footnotesize, inner sep=2pt, minimum height=8mm},
  ucaseE/.style  ={ellipse, draw=amTeal, fill=amTealF, line width=0.4pt,
                    align=center, font=\footnotesize, inner sep=2pt, minimum height=8mm},
  ucaseX/.style  ={ellipse, draw=amAmber, fill=amAmberF, line width=0.4pt,
                    align=center, font=\footnotesize, inner sep=2pt, minimum height=8mm},
  comp/.style    ={rectangle, rounded corners=2pt, draw=amBlue, fill=amBlueF,
                    line width=0.4pt, align=center, font=\footnotesize, inner sep=4pt},
  compT/.style   ={rectangle, rounded corners=2pt, draw=amTeal, fill=amTealF,
                    line width=0.4pt, align=center, font=\footnotesize\sffamily, inner sep=3pt},
  compA/.style   ={rectangle, rounded corners=2pt, draw=amAmber, fill=amAmberF,
                    line width=0.4pt, align=center, font=\footnotesize, inner sep=4pt},
  compG/.style   ={rectangle, rounded corners=2pt, draw=amGray, fill=amGrayF,
                    line width=0.4pt, align=center, font=\footnotesize, inner sep=4pt},
  compP/.style   ={rectangle, rounded corners=2pt, draw=amPurple, fill=white,
                    line width=0.4pt, align=center, font=\footnotesize, inner sep=4pt},
  compR/.style   ={rectangle, rounded corners=2pt, draw=amRed, fill=amRedF,
                    line width=0.4pt, align=center, font=\footnotesize, inner sep=4pt},
  lifeline/.style={draw=amGray, line width=0.4pt, dash pattern=on 2pt off 2pt},
  msg/.style     ={-{Stealth[length=2mm]}, line width=0.5pt},
  ret/.style     ={-{Stealth[length=2mm]}, line width=0.4pt, dash pattern=on 3pt off 2pt},
  dep/.style     ={-{Triangle[open, length=2.4mm, width=2.4mm]}, line width=0.4pt, dash pattern=on 3pt off 2pt},
  msglab/.style  ={font=\scriptsize, align=center, inner sep=1.5pt, fill=white},
}

\renewcommand{\arraystretch}{1.15}
\usepackage{listings}
\lstdefinestyle{ampy}{
  language=Python,
  basicstyle=\ttfamily\scriptsize,
  keywordstyle=\color{amBlue}\bfseries,
  commentstyle=\color{amTeal}\itshape,
  stringstyle=\color{amCoral},
  numberstyle=\tiny\color{amGray},
  showstringspaces=false,
  breaklines=true,
  frame=single,
  rulecolor=\color{amGray!45},
  backgroundcolor=\color{amGrayF!45},
  framesep=4pt,
  xleftmargin=6pt,
  aboveskip=7pt,
  belowskip=3pt,
  columns=fullflexible,
  keepspaces=true,
  captionpos=b,
  literate={->}{{$\to$}}2 {`}{{\textquotesingle}}1,
}
\title{\textbf{AutoMOOSE: Use Case and Logical Views of Agentic
Phase-Field Simulation Software}}

\author[1,2]{Sukriti Manna}
\author[2]{Henry Chan}
\author[1, 2]{Subramanian Sankaranarayanan}
\affil[1]{Department of Mechanical and Industrial Engineering,
University of Illinois Chicago, Chicago, IL 60607, USA}
\affil[2]{Center for Nanoscale Materials, Argonne National Laboratory,
Lemont, IL 60439, USA}
\affil[ ]{\texttt{smanna@anl.gov}, \texttt{skrssank@anl.gov}}

\date{}

\begin{document}
\maketitle
\vspace{-2.2em}

\begin{abstract}
\noindent
AutoMOOSE is an agentic software framework that converts a natural-language
request into an executed, screened, and interpreted MOOSE phase-field
simulation. Here, we deploy AutoMOOSE as a agentic software, complementing our prior work which focused on development of the agentic tool. We describe our software framework and architecture through
Use Case and logical views of the 1+5 architectural-views model, covering its
user roles, component structure, six-agent pipeline, physics plugin layer,
Model Context Protocol interface, and screening/falsification/recovery loop. Our
architecture separates physical falsification from automatic repair, so
corrected simulations remain inspectable and must be re-admitted before
acceptance. We focus on software design, extensibility, interoperability, and
reuse of the AutoMoose framework for broad utilization in multiphysics materials design problems.
\end{abstract}


\vspace{0.6em}
\begin{center}
{\large\textbf{Code metadata}}
\end{center}
\vspace{-0.4em}

\renewcommand{\arraystretch}{1.25}
\begin{longtable}{>{\raggedright\arraybackslash}p{0.44\textwidth}
                  >{\raggedright\arraybackslash}p{0.50\textwidth}}
\toprule
\textbf{Code metadata description} & \textbf{Details} \\
\midrule
\endhead
Current code version & 0.2.0 \\
Permanent link to code / repository used for this version &
  \url{https://github.com/sukritimanna/AutoMOOSE} \\
Permanent link to Reproducible Capsule &
  Zenodo archive deposited on acceptance \\
Legal code license & MIT License \\
Code versioning system used & git \\
Software code languages, tools, and services used &
  Python ($\geq$3.10), JavaScript/React (Vite);\newline
  FastAPI, Starlette/uvicorn (backend);\newline
  Model Context Protocol (MCP) \\
Compilation requirements, operating environments, dependencies &
  Python $\geq$3.10, Node.js $\geq$18;\newline
  a compiled MOOSE \texttt{phase\_field-opt} binary;\newline
  SLURM (HPC execution only);\newline
  a language-model backend (hosted or open-weights) \\
Link to developer documentation / manual &
  \url{https://automoose.readthedocs.io} \\
Support email for questions & \texttt{smanna@anl.gov} \\
\bottomrule
\end{longtable}
\renewcommand{\arraystretch}{1.15}


\section{Motivation and significance}
Phase-field simulation is a workhorse of computational materials
science~\cite{chen2002,moelans2008, manna2023understanding}, yet driving a general multiphysics
solver such as the Multiphysics
Object-Oriented Simulation Environment (MOOSE)~\cite{gaston2009,moose} requires a
hand-written input file that encodes the mesh, the field variables, the
free-energy kernels, the boundary conditions, the solver settings, and the
post-processors. Authoring a correct input file is a specialist task, and a
small error is often indistinguishable from a physically meaningful result
until the run is inspected by eye. AutoMOOSE removes that barrier: it
converts a single natural-language request into an executed and screened
MOOSE phase-field simulation, with no manual editing of solver input.

AutoMOOSE is aimed at computational materials researchers, students, and
AI-for-science workflow developers who use --- or would like to use ---
MOOSE for phase-field problems but are slowed by input authoring rather than
by the science. It addresses a practical barrier: translating a scientific
intent into a syntactically valid and physically meaningful MOOSE input
file, executing it, diagnosing failure modes, and interpreting the result
all require solver-specific expertise. Unlike a generic language-model
prompt, AutoMOOSE couples language-model reasoning to a constrained software
architecture --- physics plugins, execution control, verification agents, and
persistent run records --- so that the product is not a draft input file but
an executed, screened, and interpreted simulation. This agentic approach
mirrors recent autonomous-experimentation systems in
chemistry~\cite{boiko2023,bran2024}, applied here to phase-field simulation
and coupled to a constrained simulation architecture.

Our framework is a pipeline of six language-model agents, each owning one
epistemic role --- planning, input authoring, solver execution, run
screening, physics-invariant falsification, and interpretation. Its distinctive
feature is that the agent which \emph{detects} a physically implausible
result is separated from the module that \emph{corrects} a recoverable
failure: the Skeptic agent falsifies but never repairs, and a separate
closed-loop module acts on its verdict. The scientific evaluation of the
framework --- a pre-registered 25-task grain-growth benchmark~\cite{allen1979,fan1997}, recovery of
an activation energy from an ensemble Arrhenius analysis, and validation on
a second, conserved-dynamics domain (Fe--Cr spinodal decomposition~\cite{cahn1958}) --- is
reported in the companion article~\cite{companion}. This work is
concerned primarily with the software architecture: what the package does, who
uses it, how its source is structured, and how its parts interact at run
time.

AutoMOOSE sits between three nearby categories of software. MOOSE by itself is
a solver framework: it runs a phase-field problem but neither authors the
input file nor judges the result. Workflow engines such as Parsl~\cite{babuji2019},
FireWorks~\cite{jain2015}, and AiiDA~\cite{pizzi2016} orchestrate and
provenance-track simulation campaigns,
but they assume the inputs already exist and do not reason about the
physics. Generic language-model coding assistants can draft an input file
from a prompt, but they are not constrained by a solver-specific plugin
layer, do not execute or screen the run, and leave no auditable record.
AutoMOOSE combines these concerns: plugin-constrained language-model
authoring, solver execution and screening, physics-invariant falsification, and a
persistent run record, all reachable through an MCP interface.

We describe that architecture using a representative use case and logical views of the
1+5 architectural-views model~\cite{gorski2021,gorski2026}. The model
defines six views, but two of them apply to every software package, and we
use both. The Use Case view names the package's functions and the roles
that exercise them. The Logical view does two jobs: it shows the
source-code structure (as UML class or component diagrams) and the run-time
operation (as UML sequence diagrams)~\cite{gorski2026}. The model also
requires the views to agree: every role named in the "Use Case" view, and
every component named in the structure, must reappear on the lifelines of
the sequence diagrams. This discipline is what makes the two views worth the
effort for an agentic framework. Such systems are easy to sketch as a single
high-level block diagram; the 1+5 views instead force an account that is at
once structural and operational.

\section{Software description}

\subsection{Software architecture}
AutoMOOSE is organized as four layers. A request enters through one of two
interface surfaces --- a React frontend (\texttt{App.jsx}) for interactive
use, or a Model Context Protocol (MCP) server (\texttt{mcp\_server.py}, ten
tools) for programmatic clients --- both of which call a FastAPI backend
(\texttt{server.py}, port 8000). The backend orchestrates the six-agent
pipeline
\begin{equation}
S \;=\; f_5 \circ f_6 \circ f_4 \circ f_3 \circ f_2 \circ f_1\,(U),
\label{eq:pipeline}
\end{equation}
where $U$ is the user request, $f_1$ is the Architect (plan), $f_2$ the
Input Writer, $f_3$ the Runner, $f_4$ the Reviewer (execution screen),
$f_5$ the Visualization agent, and $f_6$ the Skeptic (falsification).
The composition is read right to left: $f_1$ is applied first and $f_5$
last, so the run-time order is
$f_1\!\to\!f_2\!\to\!f_3\!\to\!f_4\!\to\!f_6\!\to\!f_5$ --- the Skeptic
falsifies before the Visualization agent interprets --- with the recovery
branch entered only on a falsified, recoverable failure
(Section~\ref{sec:operation}).
Physics is resolved through a plugin layer (\texttt{plugin\_registry.py},
loading the \texttt{grain\_growth} and \texttt{spinodal} plugins), and the
language-model calls route through a provider-independent client
(\texttt{llm/}), keeping the framework model-agnostic. For local runs the
Runner executes the MOOSE solver directly; for HPC it submits through SLURM.

Listing~\ref{lst:pipeline} shows the pipeline as the backend realizes it: a
linear hand-off $f_1\!\to\!f_2\!\to\!f_3\!\to\!f_4\!\to\!f_6$, with early
exits when a stage cannot proceed. The backend orchestration ends after the
Skeptic's admission and the assembly of the run record; the
Visualization/Interpretation agent ($f_5$) then consumes that record in the
client application layer to render the figures and generate the
natural-language interpretation.

\begin{lstlisting}[caption={The six-agent pipeline as orchestrated by the
backend (condensed from \texttt{orchestrator.py}). Each $f_i$ enriches a
shared run record; screening ($f_4$) and falsification ($f_6$) gate the
result but never repair it.},label={lst:pipeline}]
# automoose/agents/orchestrator.py  (condensed)
def orchestrate(physics, params, backend):
    row = {"backend": backend, "provider": PROVIDER,
           "physics": physics, "params": params}
    f1_architect(physics, params)              # plan
    i = f2_input(physics, params)              # write .i  -> input_ok, input_lines
    if not i["input_ok"]:
        return _finish(row, ...)               # early exit: input rejected
    r = f3_run(physics, params)                # execute -> run_id, completed, wall_s
    if not r["completed"]:
        return _finish(row, ...)               # early exit: run failed
    f4_review(r["run_id"])                     # screen  -> review, valid, metrics
    f6_skeptic(r["run_id"], physics, params)   # falsify -> credible, falsified_by
    return _finish(row, ...)                    # f5 consumes record in the app layer
\end{lstlisting}

\subsection{Software functionalities}
\begin{figure}[t]
\centering
\includegraphics[width=0.75\linewidth]{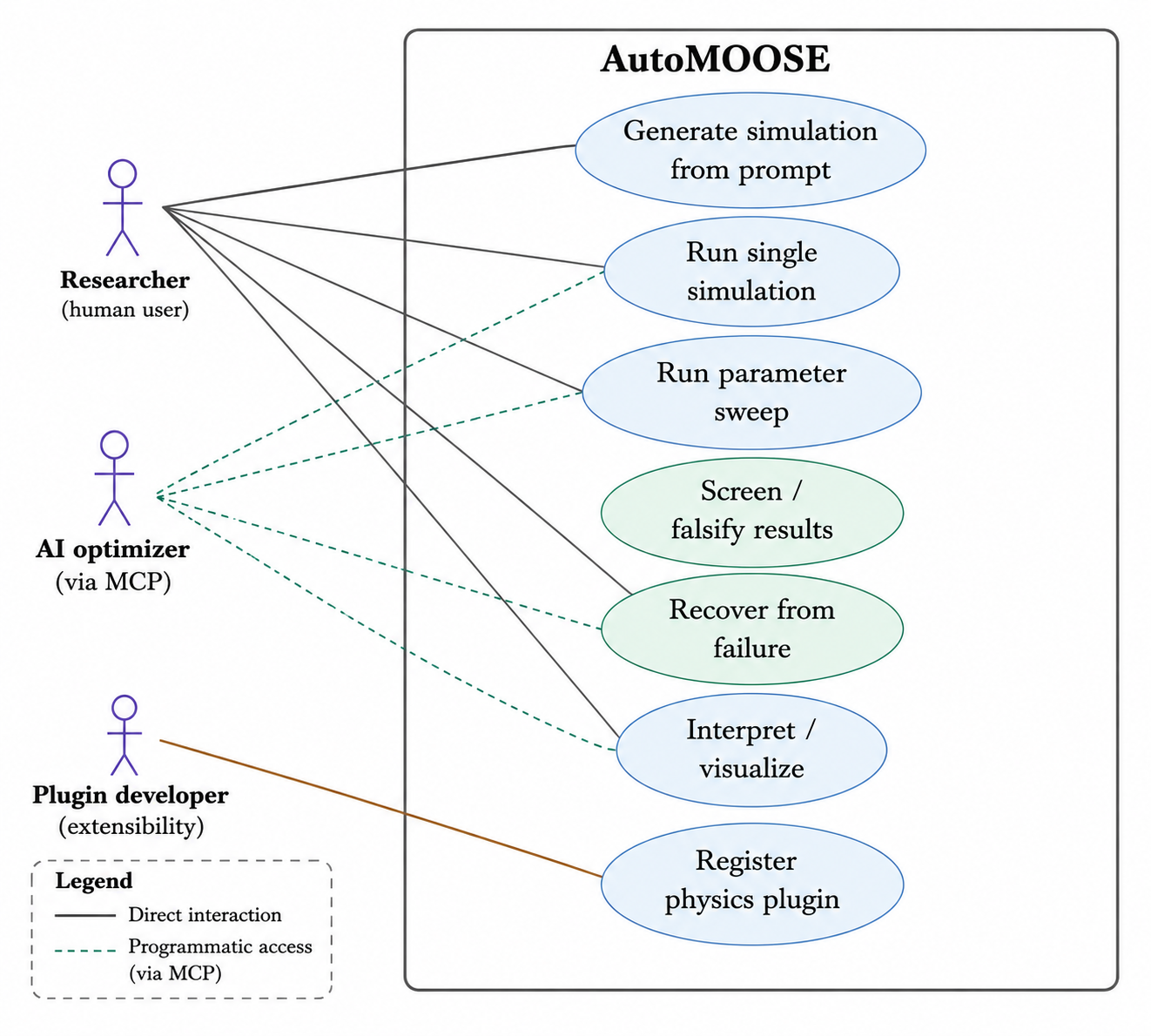}
\caption{\textbf{Use Case view.} UML use case diagram of AutoMOOSE. Three
roles (UML actors) --- Researcher, AI optimizer client, and Plugin
developer --- exercise seven functions. The Researcher reaches the full
set; the AI optimizer client reaches the run, sweep, and interpret
functions programmatically through the Model Context Protocol (MCP)
(dashed), with screening, falsification, and recovery running automatically inside them;
the Plugin developer reaches the extension function. Teal use cases are the
epistemic functions (falsification and recovery) distinctive to the
framework.}
\label{fig:usecase}
\end{figure}

With the layered architecture sketched, we now develop it through the two
applicable 1+5 views in turn --- first the Use Case view (the functions and
the roles that exercise them), then the Logical view (the source structure
and the run-time operation). AutoMOOSE offers seven functions (use cases) to
three roles (UML actors).
The roles are the \emph{Researcher}, who drives the simulation functions
interactively; an \emph{AI optimizer client}, which reaches the run, sweep,
and interpret functions programmatically through MCP --- for example, an
outer Bayesian-optimization loop; and a \emph{Plugin developer}, who
extends the physics coverage. Figure~\ref{fig:usecase} depicts the Use
Case view.

The \textit{Researcher} reaches the full set of functions: generating a simulation
from a prompt, running a single simulation, running a parameter sweep,
screening and falsifying a result, recovering from a failure, and
interpreting and visualizing output. The \textit{AI optimizer} client reaches the
run/sweep/interpret subset through MCP, with screening, falsification, and recovery
running automatically inside those functions. The \textit{Plugin developer} reaches
only the extension function, registering a physics plugin. This mirrors the
way the example real-estate package of~\cite{gorski2026} assigns its
functions to Seller and Buyer roles, here generalized to a programmatic
third actor.

\subsection{Software structure}
The Logical view first shows the source-code structure. AutoMOOSE is larger
than a single package, so we describe it at the component level --- an
abstraction the 1+5 model explicitly permits. The external MOOSE/SLURM
environment here plays the role that the external \texttt{java} package
plays in the worked example of~\cite{gorski2026}.

\begin{figure}[t]
\centering
\includegraphics[width=0.85\textwidth]{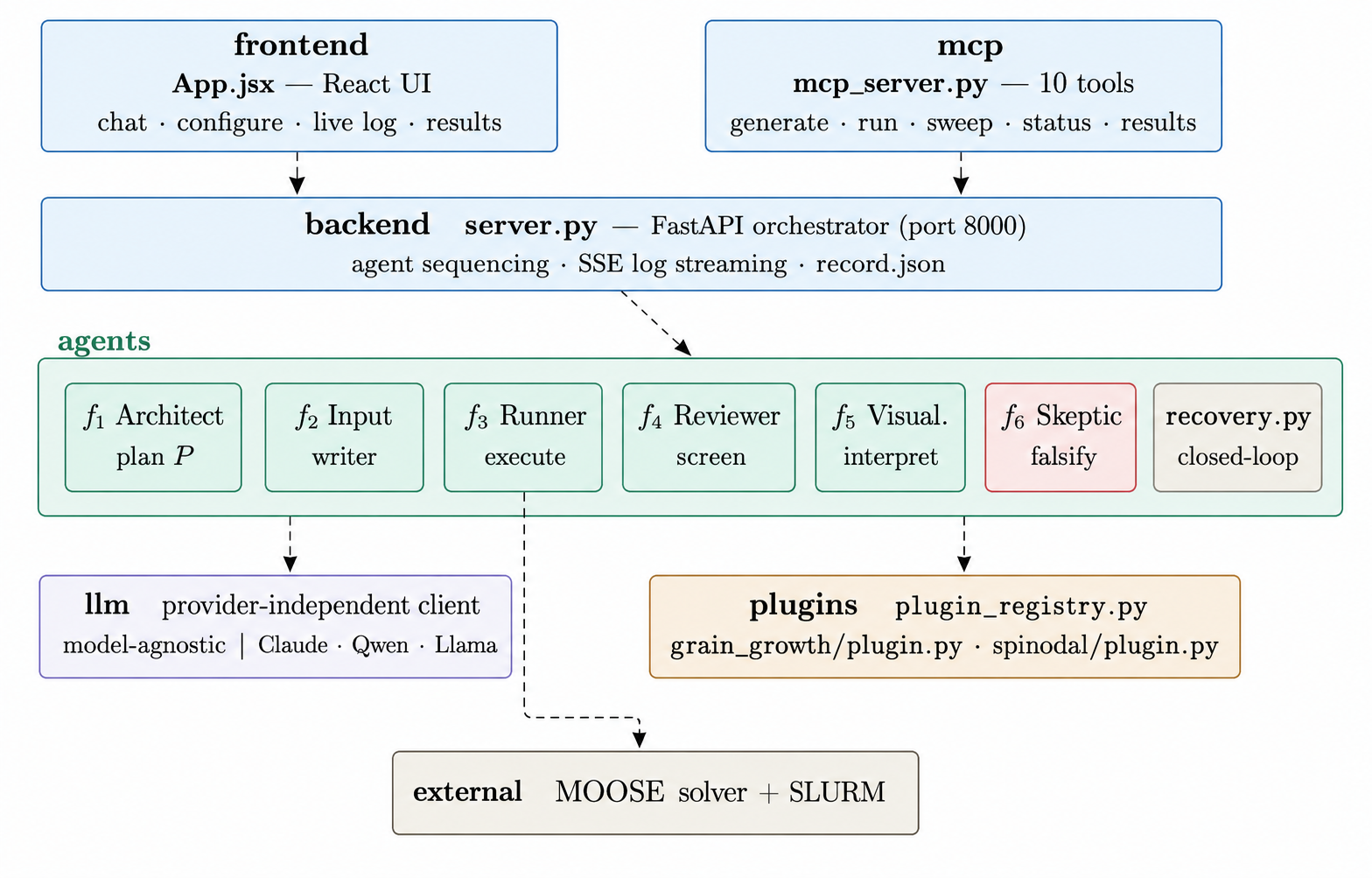}
\caption{\textbf{Logical view: structure.} UML component diagram of
AutoMOOSE. Two interface packages (\texttt{frontend}, \texttt{mcp}) depend
on the FastAPI \texttt{backend}, which orchestrates the \texttt{agents}
package; the six agents $f_1$--$f_6$ realize the pipeline
$S = f_5 \circ f_6 \circ f_4 \circ f_3 \circ f_2 \circ f_1(U)$, with the
closed-loop \texttt{recovery.py} module alongside the screening agents.
The \texttt{agents} package depends on the \texttt{plugins} package
(physics resolution) and the model-agnostic \texttt{llm} client. The
$f_3$ Runner reaches the external MOOSE/SLURM environment. Dashed open
arrows denote \texttt{<<use>>} dependencies. The
Visualization/Interpretation agent ($f_5$) appears as the sixth pipeline
role; its figures and interpretation are rendered in the frontend from the
completed run record.}
\label{fig:structure}
\end{figure}

Figure~\ref{fig:structure} reads top-down as a chain of \texttt{<<use>>}
dependencies: both interface packages depend on the backend; the backend
depends on the \texttt{agents} package; and the \texttt{agents} package
depends on the \texttt{plugins} package (physics resolution) and on the
\texttt{llm} client (model-agnostic inference). The six agents
$f_1$--$f_6$ are the classes inside the \texttt{agents} package, with the
closed-loop \texttt{recovery.py} module alongside the screening agents.
Just as the example package's \texttt{model} depends on the external
\texttt{java} package, the $f_3$ Runner depends on the external MOOSE
solver and the SLURM scheduler.

\subsection{Software operation}\label{sec:operation}
\begin{figure}[t]
\centering
\includegraphics[width=0.85\textwidth]{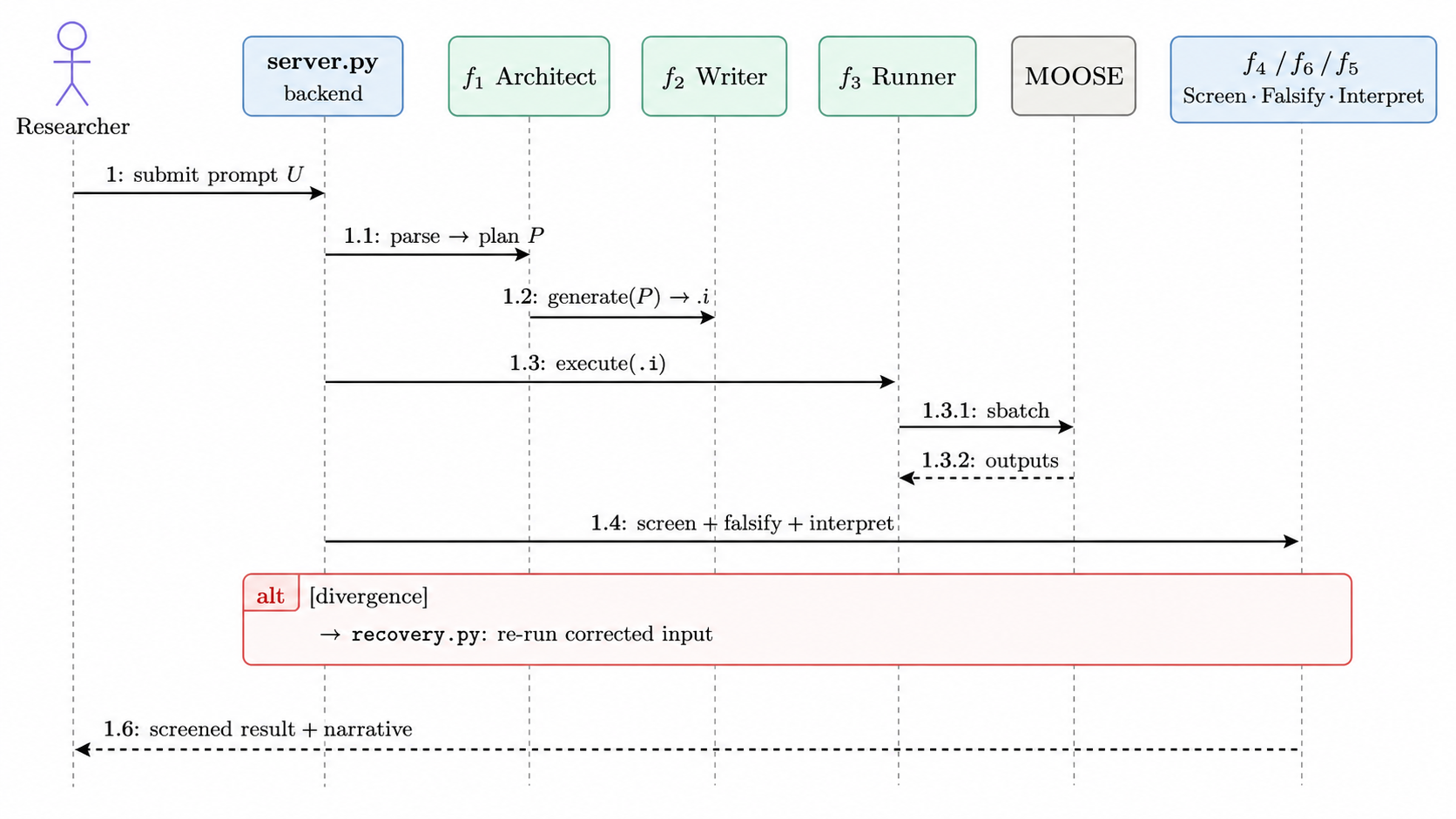}
\caption{\textbf{Logical view: operation (primary function).} UML sequence
diagram realizing the ``generate and run simulation from prompt'' use case
of Fig.~\ref{fig:usecase}. The backend sequences
$f_1\!\to\!f_2\!\to\!f_3$; the Runner launches MOOSE through
\texttt{sbatch}; and the screening--falsification--interpretation trio
($f_4$ screen, $f_6$ falsify, $f_5$ interpret) runs before the screened
result returns. The \texttt{alt} fragment is the conditional recovery
branch, entered only when the Skeptic diagnoses time-step divergence.}
\label{fig:seqmain}
\end{figure}

The Logical view also shows operation, as a sequence of method invocations
realizing each function. Following the model, we present one sequence
diagram per function and give the two most architecturally revealing here.
Figure~\ref{fig:seqmain} realizes the primary ``generate and run simulation
from prompt'' function. The backend sequences $f_1\!\to\!f_2\!\to\!f_3$; the
Runner launches MOOSE through \texttt{sbatch}; and the screening--falsification--interpretation trio ---
$f_4$ screening, $f_6$ falsification, $f_5$ interpretation --- runs before
the screened result returns to the Researcher. The \texttt{alt} fragment
is the conditional recovery branch, entered only when the Skeptic diagnoses
a recoverable time-step divergence.

\begin{figure}[t]
\centering
\includegraphics[width=0.9\textwidth]{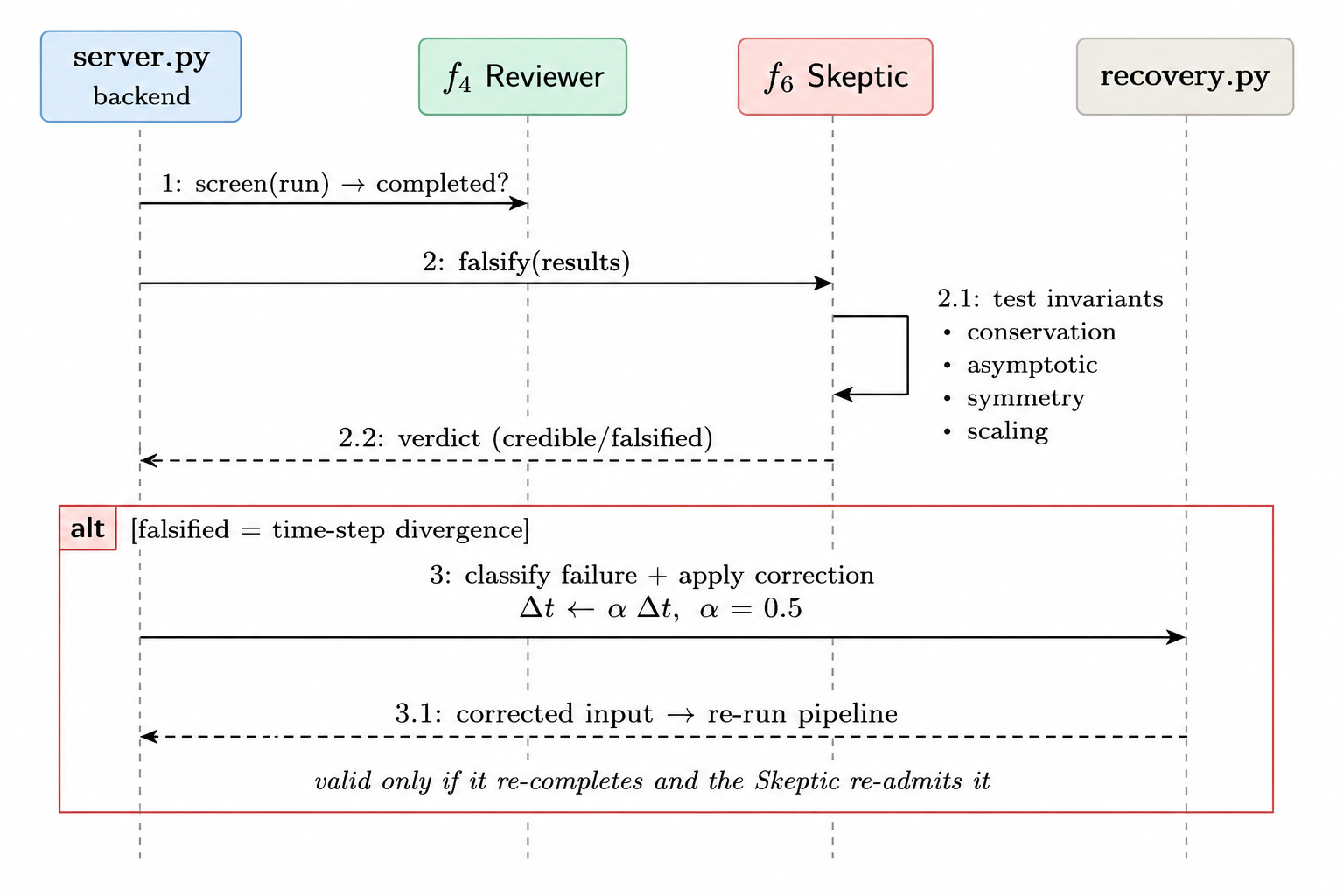}
\caption{\textbf{Logical view: operation (epistemic loop).} UML sequence
diagram realizing the ``screen/falsify'' and ``recover from failure'' use
cases of Fig.~\ref{fig:usecase}. The $f_6$ Skeptic tests the physics
invariants and returns a verdict but performs no repair; the \texttt{alt}
fragment fires only on a falsified divergence, where \texttt{recovery.py}
classifies the failure and applies the bounded time-step cutback
$\Delta t \leftarrow \alpha\,\Delta t$. A corrected run is accepted only if
it re-completes and the Skeptic re-admits it.}
\label{fig:seqloop}
\end{figure}

Figure~\ref{fig:seqloop} isolates the function with no analogue in a
conventional CRUD application and central to this framework: the
screen/falsify-and-recover loop. The $f_6$ Skeptic tests physics-grounded
invariants --- conservation, asymptotic limits, symmetry, and scaling
relations --- and returns a credible-or-falsified verdict, but it performs
no repair. The \texttt{alt} fragment fires only on a falsified divergence,
where \texttt{recovery.py} classifies the failure and applies the bounded
time-step cutback $\Delta t \leftarrow \alpha\,\Delta t$ with $\alpha=0.5$;
the corrected input is regenerated and re-run, and a corrected run is
accepted only if it independently re-completes and the Skeptic re-admits
it. Recovery is bounded in three ways: a hard cap of three attempts per
task; corrections that touch only numerical and discretization controls (the
time step, floored at a minimum value, the mesh refinement, and the
integration window) and never the physical parameters of the problem; and a
change record written for every correction. Although these edits change the
discretization, and therefore the resolution at which the simulation is
integrated, they never alter the material constants or free-energy
coefficients set by the user input. This division of labor --- detection
separated from correction --- is the structural expression of the
framework's epistemic design.

The three checks have distinct, deliberately narrow scopes.
\emph{Screening} ($f_4$) is an execution-level check --- did the run
complete and is its output present and parseable? \emph{Falsification}
($f_6$) applies plugin-specific physics invariants that can reject a result
when a declared invariant is violated; these are bounded, software-encoded
rejection criteria, not a proof of physical correctness. \emph{Recovery}
applies only predefined numerical corrections for diagnosed recoverable
failures, and is kept separate from falsification.
Listing~\ref{lst:skeptic} shows two of the grain-growth invariants as the
Skeptic implements them.

\newpage

\begin{lstlisting}[caption={Two Skeptic invariants for grain growth (condensed
from \texttt{skeptic.py}): T1 forbids spontaneous grain nucleation; T3 tests
Burke--Turnbull parabolic kinetics ($1/N$ linear in $t$, $R^2\!\ge\!0.90$).},label={lst:skeptic}]
# T1 monotonicity: grain count must not spontaneously increase
big_jumps = sum(1 for a, b in zip(n, n[1:]) if b > a + 1)
net_rise  = n[-1] - min(n)
sustained = net_rise > max(1, 0.1 * max(n))
tests["T1_monotonicity"] = {"pass": big_jumps == 0 and not sustained}

# T3 parabolic (Burke-Turnbull): 1/N linear in t, R^2 >= 0.90
pts = [(ti, 1.0 / ni) for ti, ni in zip(t, n) if ti > 0 and ni > 0]
slope, _, r2 = _linfit([p[0] for p in pts], [p[1] for p in pts])
tests["T3_parabolic"] = {"pass": r2 >= 0.90 and slope > 0, "R2": r2}
\end{lstlisting}

The recovery module turns a falsified or incomplete run into a bounded,
logged correction. Listing~\ref{lst:recovery} shows the two halves:
\texttt{classify\_failure} maps a real MOOSE log signature or a Skeptic
verdict to a failure class, and \texttt{apply\_correction} edits only
numerical and discretization controls --- each within a hard bound and
recorded as a \texttt{\{from, to, why\}} entry.

\begin{lstlisting}[caption={Bounded, logged recovery (condensed from
\texttt{recovery.py}): corrections halve the time step (floored at
\texttt{MIN\_DT0}), extend the integration window (capped at
\texttt{MAX\_END\_TIME}), or refine the mesh (capped at \texttt{MAX\_REFINE}),
never the physical parameters; the loop is capped at \texttt{MAX\_ATTEMPTS}.},label={lst:recovery}]
# automoose/agents/recovery.py  (condensed)
MAX_ATTEMPTS = 3            # generate -> run -> verify attempts per task
MIN_DT0      = 1.0          # ns; never reduce the initial timestep below this

def classify_failure(log_text, completed, skeptic_verdict):
    # map a MOOSE log signature or a Skeptic verdict to a failure class, e.g.
    # SOLVER_DIVERGENCE, KINETICS_NOT_ASYMPTOTIC, NONPHYSICAL_NUCLEATION
    ...

def apply_correction(params, diagnosis):
    """Return (new_params, change_record); pure, bounded, logged."""
    p = dict(params)                                   # never mutate the input
    cls = diagnosis["class"]
    if cls in ("SOLVER_DIVERGENCE", "NAN_DETECTED"):
        dt0 = float(params.get("dt_start", DEFAULT_DT_START))
        p["dt_start"] = max(MIN_DT0, dt0 / 2.0)        # halve, floored
    elif cls == "KINETICS_NOT_ASYMPTOTIC":
        p["end_time"] = min(MAX_END_TIME, float(params["end_time"]) * 2)
    elif cls == "NONPHYSICAL_NUCLEATION":
        p["uniform_refine"] = min(MAX_REFINE, params.get("uniform_refine", 1) + 1)
    return p, {"class": cls, "edits": _diff(params, p)}  # {from, to, why}
\end{lstlisting}

\subsection{Cross-view consistency}
The description meets the consistency that the 1+5 model
requires~\cite{gorski2026}, and it does so in three ways. \emph{First, the
roles carry across}: the Researcher who drives ``generate and run simulation
from prompt'' in the Use Case view (Fig.~\ref{fig:usecase}) is the same
actor on the lifeline of Fig.~\ref{fig:seqmain}. \emph{Second, the
components carry across}: \texttt{backend}, $f_1$--$f_6$,
\texttt{recovery.py}, and the external MOOSE are all declared in the
structure (Fig.~\ref{fig:structure}) before they appear as lifelines in
Figs.~\ref{fig:seqmain} and~\ref{fig:seqloop}; no lifeline is used that the
structure did not first introduce. \emph{Third, each use case maps to a
sequence diagram}: ``generate and run simulation from prompt'' to
Fig.~\ref{fig:seqmain}, and ``screen/falsify'' and ``recover from
failure'' to Fig.~\ref{fig:seqloop}. The remaining functions --- run sweep,
interpret, and register plugin --- would each take their own sequence
diagram to complete the Logical view.

\subsection{Physics plugins}
The two views establish what AutoMOOSE is and how it operates. Three further
concerns govern how it is reused in practice: how new physics is added, how
the framework is installed and what each run records, and how it is driven
programmatically. We take these in turn, beginning with physics extension.
\begin{sloppypar}
Physics coverage lives in a plugin layer that the agents never need to know
the internals of. A plugin is a directory under \texttt{automoose/plugins/}
containing a \texttt{plugin.py} that exposes a \texttt{PLUGIN} metadata
dictionary (label, status, parameters, and sweepable fields) together with a
module-level \texttt{generate\_input(**params)} $\to$ \texttt{str}; an optional
\texttt{parse\_results(csv\_data)} $\to$ \texttt{dict} maps solver output to metrics.
The registry (\texttt{plugin\_registry.py}) auto-discovers every such
directory at start-up, so adding physics requires no change to the agents or
the backend (Listing~\ref{lst:plugin}). Table~\ref{tab:plugins} lists the
released plugins. A stub plugin guards against accidental use on two levels:
its \texttt{generate\_input} raises \texttt{NotImplementedError}, and the
backend rejects any generate or run request for a \texttt{stub}-status plugin
with an HTTP 400 response before the pipeline starts.
\end{sloppypar}

\begin{table}[t]
\centering
\caption{Physics plugins in the released version. The two ``ready'' plugins
are the conserved- and non-conserved-dynamics domains validated in the
companion article~\cite{companion}; two further plugins are stubs in
development, included as extension templates rather than validated physics
capabilities of this release.}
\label{tab:plugins}
\small
\begin{tabular}{@{}llll@{}}
\toprule
Plugin & Physics domain & Status & Key parameters \\
\midrule
\texttt{grain\_growth}  & Allen--Cahn grain growth      & ready & \texttt{num\_grains}, \texttt{T}, \texttt{GBenergy}, \texttt{GBmob0}, \texttt{op\_num} \\
\texttt{spinodal}       & Cahn--Hilliard phase separation & ready & \texttt{c0}, \texttt{kappa}, \texttt{M}, \texttt{W}, \texttt{noise}, \texttt{end\_time} \\
\texttt{solidification} & dendritic solidification       & stub  & --- \\
\texttt{ferro}          & ferroelectric (Landau--Devonshire) & stub & --- \\
\bottomrule
\end{tabular}
\end{table}

\newpage
\begin{lstlisting}[caption={A physics plugin is a \texttt{PLUGIN} dict and a
\texttt{generate\_input} function in a per-plugin \texttt{plugin.py};
the registry auto-discovers it, with no registration call.},label={lst:plugin}]
# automoose/plugins/myphysics/plugin.py
PLUGIN = {
    "label":          "My Physics",
    "status":         "ready",          # "ready" | "stub"
    "params":         {...},            # name -> {default, range, ...}
    "sweepable":      ["T", "..."],     # parameters a sweep may vary
    "executable_key": "MOOSE_EXEC",     # env var holding the solver path
}

def generate_input(**params) -> str:
    """Return a complete MOOSE .i input file as a string."""
    ...

def parse_results(csv_data) -> dict:    # optional
    """Map MOOSE postprocessor CSV to a metrics dict."""
    ...
\end{lstlisting}

\subsection{Implementation and availability}
The second concern is getting the framework running and making each run
reproducible. AutoMOOSE installs from source. After cloning the repository, the backend is
set up with \texttt{pip install -r requirements.txt} (FastAPI, uvicorn, the
Anthropic SDK, and Pydantic; Python $\geq$3.10), and the optional React
interface with \texttt{npm install} in \texttt{frontend/} (Node $\geq$18). A
\texttt{config.env} file names the compiled MOOSE executable
(\texttt{MOOSE\_EXEC}) and the language-model backend. Because reasoning is
routed through a provider-agnostic client (\texttt{llm/}), the same pipeline
runs on different model backends --- for example Claude, Qwen, or Llama ---
by editing \texttt{config.env} rather than the code. The command
\texttt{bash start.sh} launches the FastAPI backend (port 8000) and the
frontend; the MCP server (\texttt{automoose/mcp\_server.py}) runs in stdio
mode for Claude Desktop or in SSE mode on port 8001. A new user can confirm
a working installation by setting \texttt{MOOSE\_EXEC}, launching the
backend, calling the MCP \texttt{health\_check} tool, and running the
grain-growth example of Listing~\ref{lst:mcp}; a successful install returns
a \texttt{run\_id}, a completed status, parsed coarsening metrics, and a run
record. Failures are surfaced at well-defined points: \texttt{health\_check}
catches a missing executable before any run, the Reviewer marks a run failed
when output is absent or unparseable, the Skeptic rejects a result that
violates an invariant, and \texttt{stop\_run} terminates an active job.

Every run produces a structured record that ties natural-language intent to
numerical output. The orchestrator emits, per run, the model backend and
provider, the physics and parameters, the generated input size, the run
identifier and terminal status, the wall-clock time, the Reviewer's verdict
with extracted metrics, and the Skeptic's verdict (\texttt{credible},
\texttt{falsified\_by}, and a diagnosis). Recovery actions are logged
alongside, so an automatically corrected run carries an audit trail from
prompt to accepted result. Listing~\ref{lst:record} lists the fields the
orchestrator writes for every run.

\begin{lstlisting}[caption={Fields written to each run record (from
\texttt{orchestrator.py}), grouped by the agent that produces them. The
record, not a re-query of the language model, is the unit of provenance and
reproducibility.},label={lst:record}]
record = {
    "backend": str, "provider": str, "model": str,     # language-model identity
    "physics": str, "params": dict,                    # what was requested (f1)
    "input_ok": bool, "input_lines": int,              # generated-input size (f2)
    "run_id": str, "run_status": str, "wall_s": float, # execution (f3)
    "review": str, "valid": bool, "metrics": dict,     # screening + observables (f4)
    "credible": bool, "falsified_by": list,            # falsification verdict (f6)
    "skeptic_diagnosis": str,
}
\end{lstlisting}

Because language-model outputs may vary across providers and model versions,
AutoMOOSE treats the generated MOOSE input file together with the run record
as the reproducible artifacts. Re-executing an accepted simulation therefore
does not require re-querying the language model: it requires the saved
\texttt{.i} file, the recorded parameters, and the solver environment.
Recording the software commit, plugin version, language-model identity, the
MOOSE version, a hash of the generated \texttt{.i} file, and the random seed
where the initial condition is stochastic further tightens this guarantee and
is recommended when results are to be archived.

\subsection{The MCP interface}
The third concern is driving the framework programmatically: it exposes ten
Model Context Protocol tools (Table~\ref{tab:mcp}). These cover the run,
sweep, monitoring, and
retrieval surface that an outer optimization loop needs; falsification and
recovery are not separate tools but run inside the backend pipeline that
\texttt{run\_simulation} and \texttt{run\_sweep} invoke. A sweep launches one
independent pipeline run per parameter value --- each with its own run
identifier, run record, and fresh model context --- so language-model context
does not accumulate across the sweep, and long solver logs are exposed through
\texttt{get\_log\_tail} rather than fed back in full.

\begin{table}[t]
\centering
\caption{The ten MCP tools exposed by \texttt{mcp\_server.py}. Screening,
falsification, and recovery execute inside the backend pipeline rather than
as separately callable tools.}
\label{tab:mcp}
\small
\begin{tabular}{@{}lll@{}}
\toprule
MCP tool & Function & Typical caller \\
\midrule
\texttt{health\_check}    & verify backend and MOOSE executables are found       & any \\
\texttt{list\_plugins}    & list physics plugins, parameters, and sweep fields    & any \\
\texttt{generate\_input}  & preview a MOOSE \texttt{.i} file without running       & researcher / optimizer \\
\texttt{run\_simulation}  & launch one simulation; returns a \texttt{run\_id}      & researcher / optimizer \\
\texttt{run\_sweep}       & launch a parallel sweep over a parameter list          & optimizer \\
\texttt{get\_run\_status} & poll status: pending / running / done / failed         & optimizer \\
\texttt{get\_results}     & retrieve metrics ($N(t)$, $R^2$, $\mathrm{d}N/\mathrm{d}t$, \ldots) & researcher / optimizer \\
\texttt{list\_runs}       & browse run history with optional filters               & researcher / optimizer \\
\texttt{get\_log\_tail}   & read the last $N$ lines of the solver log              & researcher \\
\texttt{stop\_run}        & terminate an active simulation                         & researcher / optimizer \\
\bottomrule
\end{tabular}
\end{table}

\newpage
\begin{lstlisting}[caption={Driving a simulation programmatically: the MCP
tools (top) and the equivalent FastAPI REST calls on port 8000
(bottom).},label={lst:mcp}]
# Via the MCP tools (e.g. from an outer optimization loop)
run_simulation(physics="grain_growth",
               params={"T": 450, "num_grains": 15, "dim": 2})   # -> run_id
get_run_status(run_id)                                          # -> "done"
get_results(run_id)                                             # -> metrics

# Equivalent over the FastAPI REST backend (port 8000)
import httpx
started = httpx.post("http://localhost:8000/run",
                     json={"physics": "grain_growth",
                           "params": {"T": 450, "num_grains": 15}}).json()
run_id  = started["run_id"]
status  = httpx.get(f"http://localhost:8000/runs/{run_id}").json()["status"]
\end{lstlisting}

\section{Illustrative example}
A minimal session exercises the primary function end-to-end. The Researcher
submits a request such as \emph{``simulate copper grain growth at 450\,K on
a 15-grain polycrystal and report the coarsening kinetics.''} The Architect
($f_1$) produces a structured simulation plan; the Input Writer ($f_2$)
emits a complete MOOSE input file; the Runner ($f_3$) submits the job
through \texttt{sbatch} and streams the solver log back over Server-Sent
Events; the Reviewer ($f_4$) confirms the run completed; the Skeptic
($f_6$) tests the result against the grain-growth invariants (grain-count
monotonicity, Burke--Turnbull parabolic scaling, and numerical integrity);
and the
Visualization agent ($f_5$) returns the fitted kinetics with a
natural-language interpretation. Each run leaves a self-contained directory
of artifacts: the generated \texttt{.i} input, the solver logs, the
postprocessor CSV files, the parsed metrics, and \texttt{record.json}.

Programmatically, the same simulation is a single MCP \texttt{run\_simulation}
call (Listing~\ref{lst:mcp}), after which \texttt{get\_run\_status} polls to
completion and \texttt{get\_results} returns the metrics (the grain-count
trajectory $N(t)$, the Burke--Turnbull~\cite{burke1952} parabolic-fit $R^2$, and the
coarsening rate). Equivalently, the headless orchestrator drives the whole
$f_1\!\to\!f_6$ loop from the command line,
\begin{verbatim}
python -m automoose.agents.orchestrator \
    --physics grain_growth --params '{"T": 450, "num_grains": 15}'
\end{verbatim}
which is how an outer optimization loop uses the framework without the user
interface. The quantitative behavior of these runs --- benchmark pass
rates, fitted activation energy, and the second-domain validation --- is
reported in~\cite{companion}.

\section{Impact and reuse potential}
The architectural contribution of AutoMOOSE is to give an agentic
scientific-computing framework an account that is both structural and
operational, rather than a single block diagram. Describing the framework
in the Use Case and Logical views makes three things explicit that a block
diagram hides. \emph{First, the roles}: the framework is usable both
interactively by a researcher and programmatically by an automated
optimizer, and the Use Case view names that second actor as a first-class
user rather than an afterthought. \emph{Second, the structure}: the
model-agnostic \texttt{llm} client and the plugin layer are separate
dependencies of the \texttt{agents} package, which is what makes the
framework portable across language-model backends and extensible to new
physics without touching the agents. \emph{Third, the operation}: the
sequence diagrams show that falsification and correction are distinct steps
with distinct owners, which is the property that lets a corrected run be
accepted only after independent re-admission.

Concretely, AutoMOOSE supports three reuse modes. It can serve as an
interactive assistant for researchers learning or running MOOSE phase-field
simulations; as a headless simulation service behind optimization,
active-learning, or parameter-sweep workflows through the MCP and REST
interfaces; and as an architectural template for wrapping other scientific
solvers with plugin-constrained agents, execution monitoring, and separated
falsification and recovery. In that third mode the same two views describe
any pipeline that turns a natural-language request into a screened and recorded artifact
through a chain of specialized agents, not only phase-field simulation: the
plugin and backend-client seams generalize directly to other solvers and
model providers. Concretely, the agent orchestration is designed to be
solver-agnostic: porting AutoMOOSE to a different solver would primarily
require replacing the plugin layer and the Runner's execution backend, while
reusing the same high-level Architect--Writer--Reviewer--Skeptic--recovery
pattern. Because a
plugin's \texttt{generate\_input} returns the solver input as an opaque
string, a target that consumes XML or JSON rather than MOOSE's \texttt{.i}
format is accommodated entirely within the plugin: the agents and backend
never parse the input deck and are unaffected by its syntax. The framework
is released under the MIT License with
documentation, so the architecture described here is directly inspectable,
reusable, and extensible through the documented plugin and MCP interfaces.

\section{Limitations and future work}
AutoMOOSE has scope limits that follow directly from its design. Its physics
coverage is bounded by the plugins currently implemented (grain growth and
spinodal decomposition); an intent that falls outside a registered plugin is
not supported. The framework assumes a working MOOSE installation and, for
HPC execution, a SLURM scheduler. Because the agents reason through a
language model, their behavior depends on the chosen backend and its
availability. Falsification is physics-specific: the Skeptic falsifies a result
only against the invariants currently defined for that physics (grain growth
and spinodal), so a domain without such invariants cannot be screened
automatically. Automatic recovery is
deliberately conservative --- it is limited to failure modes for which a
bounded correction can be justified, such as time-step divergence --- and
human inspection remains advisable when extending the framework to new
physics or producing publication-grade results. Software maturity is still
developing, though the test infrastructure is in place: the repository ships
a 25-task grain-growth evaluation set (run by \texttt{run\_evalset.py} and
scored by \texttt{score\_evalset.py}), recovery-validation scripts, and an
offline unit-test suite run under continuous integration (GitHub Actions).
The suite covers plugin discovery, \texttt{PLUGIN} schema validation, MCP
tool contracts, run-record serialization, and the physics invariants; its
contract and schema tests run against a deterministic mock model backend, so
they execute offline in CI without incurring language-model API calls.
Broadening this coverage --- for example towards end-to-end,
solver-in-the-loop integration tests --- remains ongoing.

Future work will broaden the plugin library to further MOOSE modules and make
falsification invariants part of the plugin contract through a plugin-level
\texttt{validate\_metrics} hook, moving physics-specific falsification
criteria out of the central Skeptic and into the same reuse boundary that
already controls input generation and result parsing. We also plan to extend
the falsification layer to additional classes of multiphysics simulation, add
containerized reference deployments and schema-validated agent outputs for
reproducibility, and strengthen support for offline, open-weight model
backends.

\section{Conclusions}
We have described the architecture of AutoMOOSE, an agentic framework for
phase-field simulation, the Use Case and Logical views of
the 1+5 architectural-views model. The Use Case view names seven functions
and three roles; the Logical view depicts the component structure of the
six-agent pipeline together with its external dependencies, and the
operational sequences that realize its primary and epistemic functions. The
description is cross-consistent: roles and components declared in one view
reappear in the others. The scientific validation of the framework is
reported separately~\cite{companion}; the architecture documented here is
what makes that framework portable, extensible, and inspectable.

\section*{Declaration of competing interest}
The authors declare that they have no known competing financial interests
or personal relationships that could have appeared to influence the work
reported in this paper.

\section*{Declaration of generative AI use}
The AutoMOOSE framework itself uses large language models as the reasoning
engine of its agents, as described above. The authors also used an AI
assistant during the preparation of this manuscript; all technical content,
architecture, and claims were verified by the authors against the source
code and the companion article.

\section*{Code and data availability}
AutoMOOSE is openly available at \url{https://github.com/sukritimanna/AutoMOOSE}
under the MIT License, with documentation at
\url{https://automoose.readthedocs.io}. The version described here is
AutoMOOSE \texttt{v0.2.0}, corresponding to Git commit \texttt{d92239c} on the
\texttt{main} branch. A versioned Zenodo archive with a citable DOI will be
provided for the accepted version. No experimental datasets were
generated for this software article; the benchmark data underlying the
companion study~\cite{companion} are described there.

\section*{CRediT authorship contribution statement}
\begin{sloppypar}
\textbf{Sukriti Manna:} Conceptualization, Methodology, Software,
Validation, Writing --- original draft.
\textbf{Henry Chan:} Writing --- review.
\textbf{Subramanian Sankaranarayanan:}
Funding acquisition, Writing --- review.
\end{sloppypar}

\section*{Acknowledgements}
This work was supported by the DOE Office of Science, Basic
Energy Sciences under the AI-Pathfinder project.
Work at the Center for Nanoscale Materials, a U.S.\ DOE Office
of Science User Facility, was supported under Contract
No.\ DE-AC02-06CH11357.
This work used NERSC, supported by the DOE Office of Science
under Contract No.\ DE-AC02-05CH11231.
We acknowledge LCRC computing facilities at Argonne.


\bibliographystyle{unsrtnat}
\bibliography{references}

\end{document}